\documentclass[aps,prl,twocolumn,showpacs,superscriptaddress,groupedaddress]{revtex4}  
\usepackage{graphicx}  
\usepackage{dcolumn}   
\usepackage{bm}        
\usepackage{amssymb}   

	\usepackage{amsmath}
	\usepackage{makeidx}
	\usepackage{amsfonts}
	\usepackage[ansinew]{inputenc}
	\usepackage[usenames,dvipsnames]{pstricks}
	\usepackage{subfigure}
	\usepackage{epsfig}
	\usepackage{float}
	\usepackage{pst-grad} 
	\usepackage{pst-plot} 
	\usepackage[colorlinks,hyperindex]{hyperref}
	\hypersetup
	{
		colorlinks,%
		citecolor=black,%
		linkcolor=black,%
		urlcolor=black,%
	}

\makeindex
\begin{document}
	\author{Arash Khazraie}
	\author{Kateryna Foyevtsova}
	\author{Ilya Elfimov}
	\author{George A. Sawatzky}
	\affiliation{Department of Physics $\&$ Astronomy, University of British Columbia, Vancouver, British Columbia V6T 1Z1, Canada}	
	\affiliation{Quantum Matter Institute, University of British Columbia, Vancouver, British Columbia V6T 1Z4, Canada}
\title{Bond versus charge disproportionation in the bismuth perovskites}

\date{\today}

\begin{abstract}
We develop a theory describing a parameter based phase diagram to be associated with materials incorporating skipped valence ions\cite{Varma}. We use a recently developed tight-binding approach for the bismuthates to study the phase diagram exhibiting the crossover from a bond disproportionated (BD) to a charge disproportionated (CD) system in addition to the presence of a new metallic phase. We argue that three parameters determine the underlying physics of the BD-CD crossover when electron correlation effects are small: the hybridization between O-2$p_{\sigma}$ and Bi-6$s$ orbitals ($t_{sp\sigma}$), the charge-transfer energy between Bi-6$s$ and O-a$_{1g}$ molecular orbitals ($\Delta$), and the width of the oxygen sublattice band ($W$). In the BD system, we estimate an effective attractive interaction $U$ between holes on the same O-a$_{1g}$ molecular orbital. Although here we concentrate on the example of the bismuthates, the basic ideas can be directly transferred to other perovskites with negative charge-transfer energy, like ReNiO$_{3}$ (Re: rare-earth element), Ca(Sr)FeO$_{3}$, CsTIF$_{3}$ and CsTlCl$_{3}$.

\end{abstract}

\maketitle

A common feature of the classes of materials involving skipped valence elements \cite{Varma} that we are considering here is that the ground state crystal structure consists of an alternation of long and short metal--oxygen bond lengths with an unaltered coordination number. This is commonly encountered in the perovskite structures with the chemical formula of $AB$O$_{3}$ with skipped valence $B$ site ions which are identical and octahedrally coordinated with oxygen in the cubic phase. The formal valence of the $B$ site cation is given by V($B$) = 6 $-$ V($A$), assuming the customary closed shell O$^{2-}$ electronic configuration. This leads us to the very unusual valence of 4+ for Bi in $A$BiO$_{3}$\cite{Sleight,Cava,Kazakov}($A$ = Sr, Ba) with a half filled 6$s$ shell or an also rather uncommon valence of Ni$^{3+}$(3$d^{7}$) in the rare-earth ($Re$) Nickelates $Re$NiO$_{3}$ \cite{Mizokawa, Johnston,Lau,Park, Green} or Fe$^{4+}$(3$d^{4}$) in Ca(Sr)FeO$_{3}$\cite{Takano, Bocquet}. 

Two alternative, although not orthogonal, concepts have been used to describe how skipped valence systems adjust to solve this problem. The most common approach is to assume that oxygens remain closed shell and that all the charge action is then on the cation. In this approach the Bi$^{4+}$ with the Bi-6$s$ open shell orbital disproportionates \cite{cox,cox2,Rice,Mattheiss82,Mattheiss83,Varma,Hase}, forming an empty and full (Bi$^{5+}$,Bi$^{3+}$) 6$s$ shell. This would obviously also strongly affect the Bi-O bond lengths which would be considerably shorter for the 5+ configuration than the 3+ one based on known compounds with corresponding formal valencies\cite{Sleight}. This sort of behavior of valence skipping elements is often accounted for by the concept of an attractive Hubbard-$U$ type model\cite{Varma,Rice} which still leaves open the question regarding how this interaction could become attractive since electron-phonon coupling in BaBiO$_{3}$ is generally found to be very weak from local-density approximation (LDA) calculations\cite{Hamada, Shirai, Liechtenstein, Kunc, Meregalli, Bazhirov}. 

Another proposed starting point is that of a negative-charge-transfer gap, meaning that the first ionization states keep the cation at its most common valence in oxides, which is 3+ for Bi, and removing electrons from the O-2$p$ states to form a charge neutral system\cite{Foyevtsova, Plumb, Balandeh, Khazraie}. In this approach the energy of the formal valence configurations of Bi$^{4+}$O$^{2-}$ would convert to Bi$^{3+}\underline{L}$ where $\underline{L}$ refers to a hole in the O-2$p$ band. In this case there would be on average two holes per oxygen octahedron. The energy can be further lowered in this negative charge-transfer gap system by forming a Peierls-like \cite{Peierls} bond-disproportionated state, because of the cooperative formation of double covalent bonds that form octahedra of oxygens with alternating short and long bond lengths with the central Bi. Here, the Bi-6$s$ charge will be only slightly different for the two sites and all the oxygens remain identical. Therefore, little change in the average charge density is needed. This is evident from the density functional theory (DFT) calculations of Ref.~\cite{Liechtenstein, Foyevtsova} where small charge-disproportionation of $\pm$0.15e was found inside the Bi muffin-tin spheres. In addition, spectroscopic measurements of Ref.~\onlinecite{Hair, Orchard, Wertheim} found little difference in the Bi valence shell occupations.

 The situation in the 3$d$ transition metal oxides is also similar \cite{Mizokawa, Johnston,Lau,Park, Green}, but with large differences regarding the orbitals that are involved and the fact that strong electron correlation effects prevail \cite{Zannen}. However, the basic physics remains the same upon including the correlation and the Hunds' rule physics as well as $d$ symmetry rather than $s$ symmetry orbitals in the covalent bonds.

In the end, the two starting points which we could name as charge disproportionated (CD) and bond disproportionated (BD) respectively yield the same result of a disproportionated state with the same symmetry and lattice deformation types of long and short bonds, and even same spin states for the case of the 3$d$ transition metal candidates. Therefore, the choice of the approach is not so important for describing the qualitative nature of the ground state. However, the nature of the states involved in the low energy scale properties is quite different which is important in describing the excited states of the systems. For example in the BD picture it is quite natural to obtain solutions in which two holes cooperatively bond to the short bond length Bi forming a lattice of bipolarons with an attractive interaction resulting from the electron-phonon coupling. This is important for the potential description of superconductivity in the hole doped systems as in Ba$_{1-x}$K$_{x}$BiO$_{3}$\cite{Cava}.

In this paper, we demonstrate that there is quite a natural $\it{crossover}$ between a bond and a charge disproportionated regime. In the former case, the two holes' wave function is mostly localized on the O-2$p$ molecular orbitals of a$_{1g}$ symmetry in the collapsed octahedron while in the latter case the holes' wave function is mostly localized on the Bi-6$s$ orbital of the collapsed octahedron. We argue that only three parameters determine the underlying physics of this crossover: $t_{sp\sigma}$, the strength of the hybridization between O-$2p_{\sigma}$ and Bi-$6s$ orbitals, $\Delta = \epsilon(\text{Bi-}6s) - \epsilon(\text{O-}a_{1g}$), the charge-transfer energy, and $W$, the width of the oxygen sublattice band. In addition, we show that within a certain parameter range a metallic phase appears in which the holes are in delocalized O-2$p$ bands mainly of O-e$_{g}$ symmetry which do not hybridize with the Bi-6$s$. 

\begin{figure}[t]
\begin{center}
\includegraphics[width=0.55\textwidth]{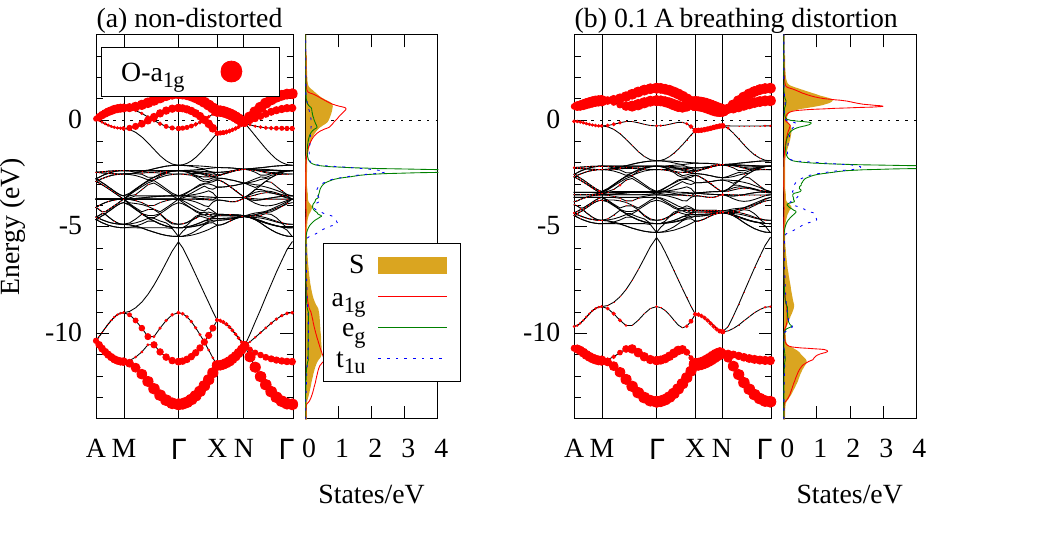} 
\caption{The band structures and projected density of states of the tight-binding model for (a) non-distorted lattice and (b) 0.1 \AA~breathing distorted lattice. Oxygen molecular orbital projections are made onto the compressed octahedron and the Bi-$6s$ orbital projection is for the Bi atom located inside the compressed octahedron. The red-colored fat bands represent the contribution of the O-a$_{1g}$ molecular orbital of the compressed octahedron. The Fermi level is marked with a horizontal dashed black line.}
\label{Fig:TB3D.pdf}
\end{center}
\end{figure}

\begin{figure}[h]
\begin{center}
\includegraphics[width=0.47\textwidth]{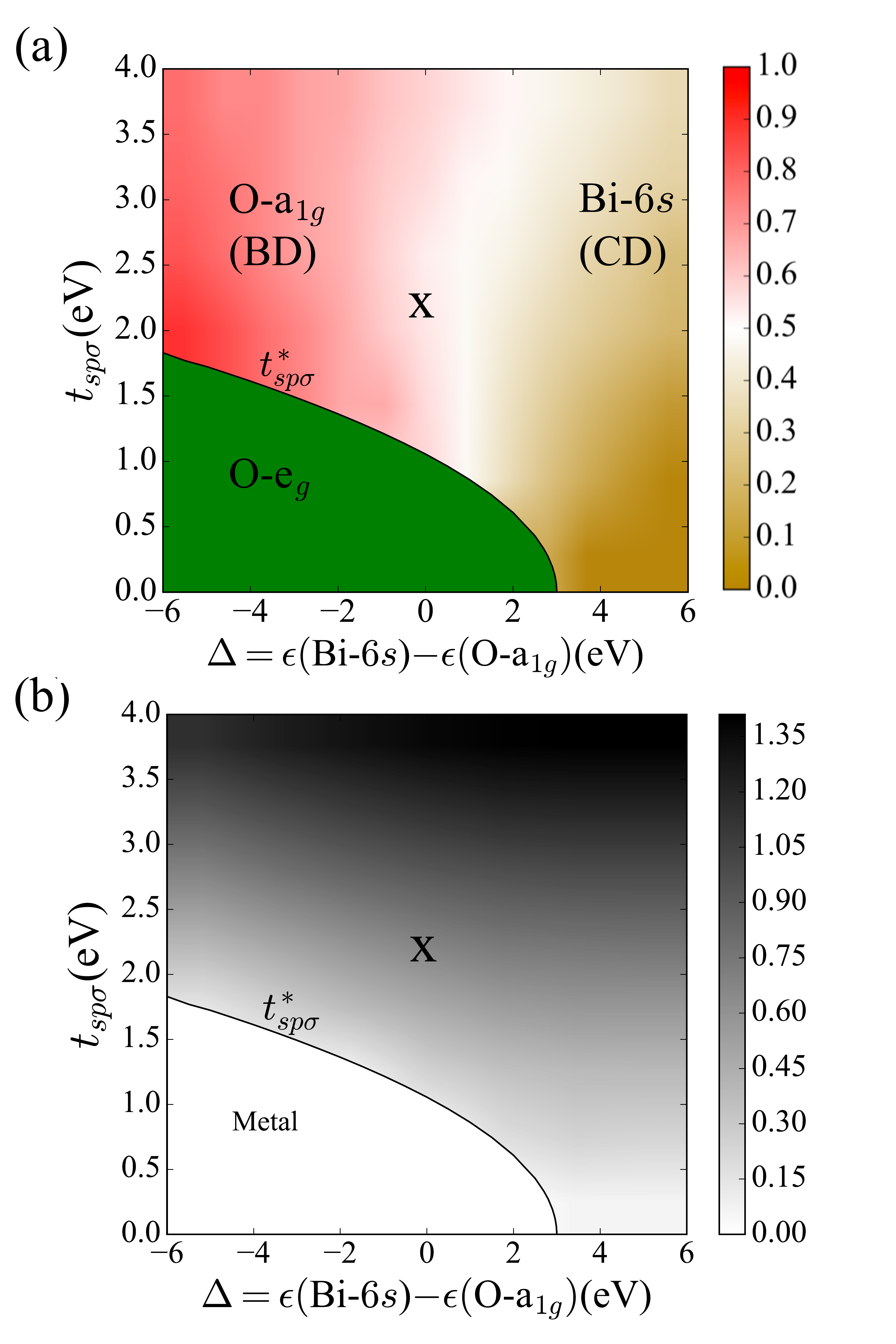} 
\caption{(a) The phase diagram representing the dominant character of the conduction band for the experimental breathing distorted lattice obtained from the tight-binding model as a function of the charge-transfer energy, $\Delta$ and t$_{sp\sigma}$. Red(yellow) represents O-a$_{1g}$(Bi-6$s$) character in the BD(CD) regime respectively. Symbol x marks the parameters relevant for $A$BiO$_{3}$ \cite{Khazraie}. (b) The charge gap in eV calculated from the tight-binding model for the breathing distorted lattice. An insulator-to-metal transition is obtained at a critical value of $t^{*}_{sp\sigma}$ when holes transition into the O-e$_{g}$ molecular orbitals.}
\label{Fig:Fig2.pdf}
\end{center}
\end{figure}

To study the BD-CD crossover in the bismuthates, we use the tight-binding (TB) model derived in Ref. \onlinecite{Khazraie} which consists of one Bi-6$s$ and nine O-$2p$ orbitals per formula unit with two nearest-neighbor inter-site O-2$p$ hopping integrals, $t_{pp\sigma}$ and $t_{pp\pi}$, and the Bi-6$s$--O-$2p_{\sigma}$ hopping integral, $t_{sp\sigma}$. This simple TB model can describe well the changes in the electronic structure such as the opening of the charge gap due to the breathing distortion observed in previous DFT studies \cite{Mattheiss82, Mattheiss83, Foyevtsova}, as well as the formation of molecular orbitals on the collapsed octahedra as shown in Fig.~\ref{Fig:TB3D.pdf}. Here for simplicity, we have neglected the tilting distortions of the oxygen octahedra, and consider a four formula unit supercell lattice with an experimental breathing distortion of 0.1 \AA\cite{Kazakov}. From the projected density of states [see Fig.~\ref{Fig:TB3D.pdf}(a)] we can see that the Bi-$6s$ band is very broad, extending from +1 eV above the Fermi level down to $-$12 eV below. This is due to the strong hybridization between Bi-$6s$ and O-a$_{1g}$ states. After introducing the breathing distortion, holes spatially condense onto O-a$_{1g}$ molecular orbitals of the collapsed BiO$_{6}$ octahedra which is represented by the red-colored fat bands in Fig.~\ref{Fig:TB3D.pdf}(b).
 
A BD-CD crossover can now be obtained by scanning regions of strong and weak hybridization and varying the charge-transfer energy (i.e. by varying Bi-6$s$ on-site energy) in a breathing distorted lattice as shown in Fig.~\ref{Fig:Fig2.pdf}(a). Here, x marks the parameters relevant for $A$BiO$_{3}$ \cite{Khazraie}. We find that the holes' character crosses over from O-a$_{1g}$ (red) to Bi-$6s$(yellow) as we go from a negative to positive charge-transfer energy. However, it is important to note that since there is no symmetry change in moving from a BD to a CD system, there is no clear boundary but rather a gradual crossover that can be defined when there is equal Bi-6$s$ and O-a$_{1g}$ character in the conduction band. Also since each octahedron has on average two holes, this average density does not change as we go from CD to a BD state but only the fluctuations strongly diminish from the average which is similar to a Peierls like transition. In addition to the BD-CD crossover we find that at a critical hybridization $t^{*}_{sp\sigma}$ for a given $\Delta$, the two holes per octahedron transition into non-bonding purely O-2$p$ states of e$_{g}$ molecular orbital symmetry located at about 2 eV below the Fermi level[see Fig.~\ref{Fig:TB3D.pdf}(b)]. 

 In the BD regime two hole states on a collapsed oxygen octahedron strongly bond with the central Bi-6$s$ orbital, which results in a lowering of the system's total energy relative to when the holes are well separated in different octahedra. This quite naturally results in tendencies to bipolaron formation and leads to an effective ``molecular"  \cite{Khazraie} rather than on-site attractive interaction $U$ which can result in superconductivity. To estimate the value of this attractive $U$, we calculate the energy of a system with two spatially well separated holes in a$_{1g}$ molecular orbitals by including the polaronic effects \cite{Su, Moeller1, Moeller2} resulting from the increase of $T_{sp\sigma} = \sqrt{6} t_{sp\sigma}$ molecular orbital hopping integral due to electron-phonon interaction and compare that with the energy of two holes of opposite spin on the same a$_{1g}$ molecular orbital. The polaronic lowering of the single hole energy due to the decrease in the Bi$-$O bond length can be estimated as $\delta T_{sp\sigma} = \sqrt{6}\delta t_{sp\sigma}$ and the total energy lowering for two of these well separated holes is twice this i.e. 2$\delta T_{sp\sigma}$. Recalling that $t_{sp} \approx 1/(d_{0}+\delta d)^{2}$ \cite{Froyen}, for small $\delta d$ , $\delta t_{sp}$ is proportional to $-\delta d$. Now, if there are two holes on the same a$_{1g}$ molecular orbital the change in the bond length will be twice as large in linear response theory and so $\delta t_{sp}$ will be twice as large and the total energy lowering will be 4$\delta T_{sp}$. So the energy difference between two holes on the same a$_{1g}$ orbital and two times one hole per a$_{1g}$ orbital will be $-2\delta T_{sp}$. Given the breathing distortion of 0.1~\AA~we calculate an effective attractive interaction $U_{eff} = -2\delta T_{sp\sigma}=-2\sqrt{6}(2.37-2.15)\approx$ $-1.1$ eV between two holes on an O-a$_{1g}$ molecular orbital. 

\begin{figure}[t]
\begin{center}
\includegraphics[width=0.45\textwidth]{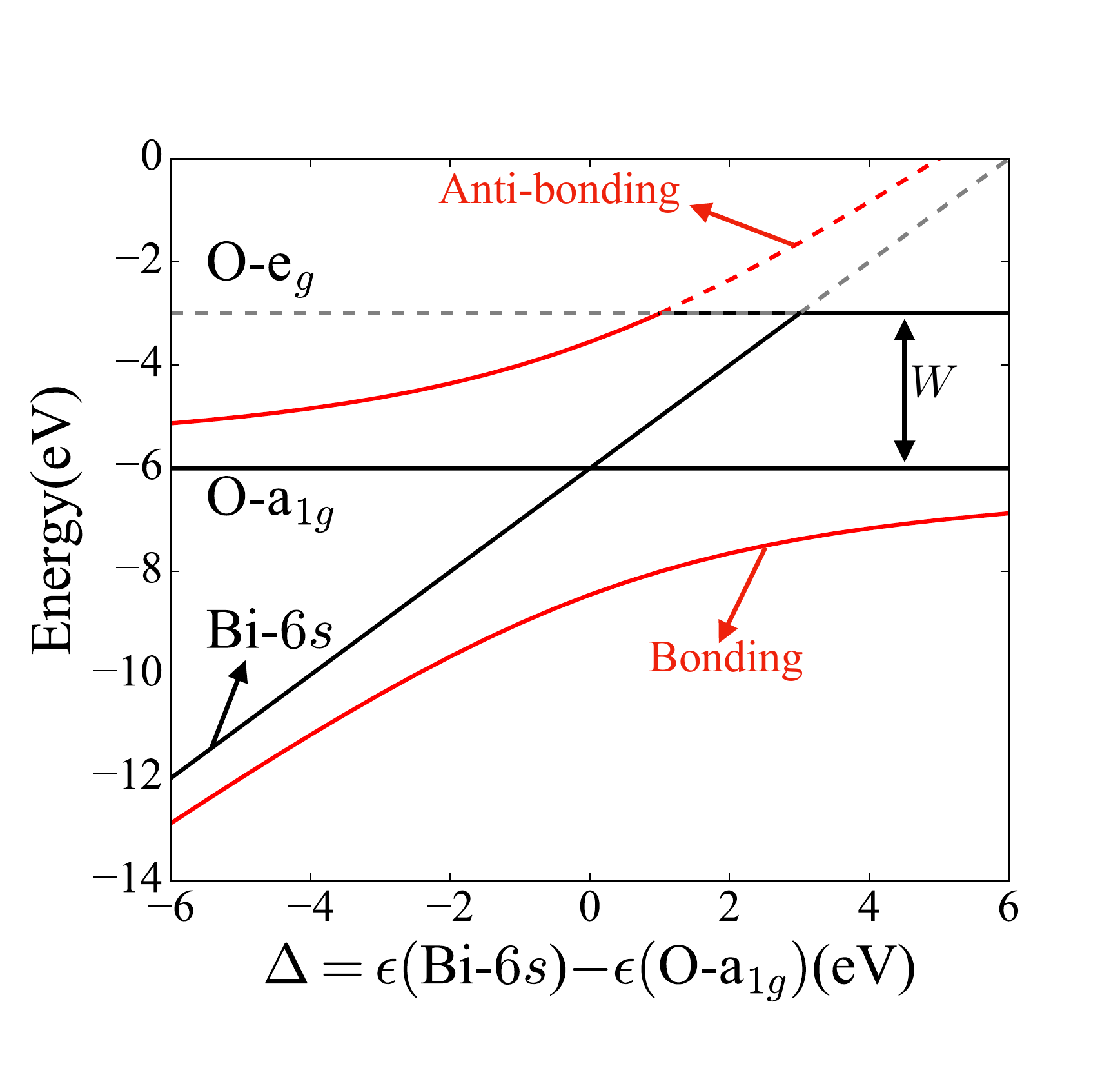} 
\caption{The energy level diagram of an oxygen octahedron calculated within the three-band model at $t_{sp\sigma} = 0$(in black), and $t_{sp\sigma}= 1$ eV(in red) as a function of the charge-transfer energy $\Delta$. The character of the conduction band (dashed line) transitions from O-e$_{g}$ to anti-bonding Bi-$6s$--O-a$_{1g}$ band at a critical value of $t_{sp\sigma}$ and charge-transfer energy ($\Delta$).}
\label{Fig:Fig3.pdf}
\end{center}
\end{figure}

Let us now calculate the charge gap in our TB model for the breathing distorted lattice and above considered ranges of $t_{sp\sigma}$ and $\Delta$. We find that the distorted structure has a gap for all values of $\Delta$ and hopping integrals larger than $t^{*}_{sp\sigma}$[see Fig.~\ref{Fig:Fig2.pdf}(b)]. However, for weaker hybridization when holes cross over into O-e$_{g}$ orbitals the system transitions into a metallic state. This insulator-to-metal transition can be understood by noting that the Peierls-like breathing distortion of the oxygen octahedra does not open a gap in the O-e$_{g}$ states, because this distortion is not coupled to the O-e$_{g}$ states due to symmetry. As a result, the system can not lower its energy by adopting the breathing distortion and therefore will stay in the metallic non-disproportionated state. 

Finally, we show that this phase transition can simply be described within a three-band model consisting of only O-a$_{1g}$, O-e$_{g}$, and Bi-6$s$ orbitals with the hybridization $T_{sp\sigma} = \sqrt{6}t_{sp\sigma}$. This hybridization couples only the O-a$_{1g}$ and Bi-6$s$ orbitals, since the O-e$_{g}$ orbital does not hybridize with either Bi-6$s$ or O-a$_{1g}$ due to symmetry. We find that at zero hybridization, O-e$_{g}$(O-a$_{1g}$) states are at the top(bottom) of the oxygen band which results in the conduction band (represented by a dashed line) being mainly of O-e$_{g}$ character[see Fig. \ref{Fig:Fig3.pdf}]. 
However, after Bi-6$s$--O-a$_{1g}$ hybridization is turned on, at a certain value of the charge transfer-energy, the anti-bonding Bi-6$s$--O-a$_{1g}$ band is pushed above the O-e$_{g}$ states. We can realize this transition for a fixed hybridization ($t_{sp\sigma}$ = 1 eV) by going from a negative to positive charge-transfer energy as shown in Fig. \ref{Fig:Fig3.pdf}. It can be seen that the conduction band changes character from O-e$_{g}$ to anti-bonding Bi-6$s$--O-a$_{1g}$ at $\Delta \approx$ 1 eV. We find that the boundary of this transition is described by a function of $\Delta$ and $W$:
\begin{equation}
t^{*}_{sp\sigma}=\frac{\sqrt{W\Delta +W^{2}}}{\sqrt{6}}.
\end{equation}
where $W$=$\epsilon$(O-e$_{g}$)$-\epsilon$(O-a$_{1g}$) is the width of the oxygen band.

In summary, we have studied the orbital character of the holes in the bismuthates and have shown that it depends strongly on the strength of Bi-6$s$ and O-$2p_{\sigma}$ hybridization, $t_{sp\sigma}$, and the charge-transfer energy, $\Delta$. We have demonstrated that there exists a natural crossover between a bond and a charge disproportionated system with holes preferring to occupy Bi-6$s$ or O-a$_{1g}$ molecular orbitals. In the BD regime, an effective attractive ``molecular" orbital interaction between two holes residing on the collapsed oxygen octahedra is estimated as $U_{eff}$ = $-$1.1 eV, via considering electron-phonon coupling effects through changes in $t_{sp\sigma}$. We have further shown that holes can transition into non-bonding O-e$_{g}$ molecular orbitals which is accompanied by an insulator-to-metal transition. One might expect that similar cross overs can occur in other perovskites with localized molecular orbital symmetries other than the bismuthates.

This work was supported by Natural Sciences and Engineering Research Council (NSERC) for Canada, Canadian Institute for Advanced Research (CIFAR), and the Max Planck-UBC Stewart Blusson Quantum Matter Institute.

\end{document}